# Is Carbon Cycling Keeping Pace with Increases in Atmospheric Carbon Dioxide?

Benjamin Helyer[a] and Michael Courtney[b]

BTG Research; 9574 Simon Lebleu Road, Lake Charles, LA, 70607, USA

[a] benjaminhelyer@tamu.edu

[b] Michael_Courtney@alum.mit.edu

## Abstract

Carbon dioxide ($CO_2$) increase has been well documented, and global net primary production is of importance to a variety of ecological topics. Since $CO_2$ increases primary production in laboratory experiments, the global effects of increasing $CO_2$ on global primary production are of interest in both climate science and ecology. Various studies have considered increases in primary production over different regions and time scales, but the global effects of increased atmospheric $CO_2$ and primary production remain unquantified. This study aims to compare these two variables globally to assist in determining the potential for increases in primary production to contribute to carbon sequestration, possibly slowing increases in atmospheric $CO_2$ resulting from fossil fuel emissions. Monthly $CO_2$ concentration data from 1985 through 2015 in distinct latitude bands (every 10º) was retrieved from NOAA's Earth System Research Laboratory for a total of 18 datasets. As a proxy to quantify net primary production, the magnitude of annual $CO_2$ cycling was computed for each dataset through Fourier analysis. Relative increases were then calculated for both $CO_2$ increase and amplitude increase to compare the pace of carbon cycling with the pace of increases in atmospheric $CO_2$. Globally, the increase in primary production determined by using annual $CO_2$ cycling as a proxy was 3.75% (±1.82%) per decade over the time interval studied. In contrast, the measured increase in $CO_2$ abundance was 4.75% (± 0.02%) per decade. Increases in carbon cycling appear to be slightly smaller than increases in $CO_2$ globally and in both the Northern and Southern Hemispheres, but the uncertainties in the estimate for increases in carbon cycling are too large to draw that conclusion with statistical confidence.

**Introduction**

The rise in atmospheric carbon dioxide ($CO_2$) concentration has been well documented for over 60 years (Keeling et al., 2012). The result that increased $CO_2$ causes increased productivity for producers has been reported in numerous studies using a variety of methods, ranging from short-term samples from oceanic environments (Hare et al., 2007; Hein & Sand-Jensen, 1997) to longer-term investigations in terrestrial environments (Ainsworth & Long, 2005; DeLucia et al., 1999; Fernández-Martínez 2017; Schlesinger et al., 2009). For individual plants, it has been suggested that an increase of $CO_2$ increases water use efficiency of stomata as well as their rate of photosynthesis (Drake, Gonzalez-Meler, & Long, 1997).

However, observed regional and global increases in net primary production (NPP) may not be directly due to the stimulation of production by increased $CO_2$. One study found that terrestrial net primary production (NPP) increased an average of 6% globally from 1982 to 1999 but attributed the result primarily to climatic changes, such as increases in temperature and solar radiation. To support this attribution, an example of an increase in tropical NPP was noted in the study. Because tropical regions have not been shown to be large carbon sinks, the study contended that production in these areas increased due to factors other than direct stimulation by increased $CO_2$ (Nemani et al., 2003).

A 2010 study, using a method similar to that of Nemani et. al., found that global terrestrial NPP decreased from 2000 to 2009, despite continued increases in temperature and $CO_2$ (Zhao & Running, 2010). The change was attributed to reduced rainfall in the tropics and Southern Hemisphere, and despite global decrease, NPP continued to increase in the Northern Hemisphere. Because temperature is not believed to be the leading limitation on the length of the growing season in the Southern Hemisphere, the authors suggested that any increase from rising temperatures on NPP could be more easily offset by other factors, such as drought, in this hemisphere. Similar reasoning was used for increases in the Northern Hemisphere, where rising temperatures could lengthen growing seasons sufficiently to counterbalance simultaneous drying. In summary, the result of net global decrease in NPP found in this 2010 study could be caused by decreases in the Southern Hemisphere, indicating the Northern Hemisphere has still continued to act as a carbon sink. Thus, the above-mentioned study does not negate the possibility that the magnitude of annual $CO_2$ oscillation is a good choice of proxy for NPP in the Northern Hemisphere. However, because data are readily available, the present study also analyzes carbon cycling in the Southern Hemisphere, despite the possible effects of other factors at work in this hemisphere.

Although $CO_2$ increases may or may not have directly caused changes in NPP over the past two decades, terrestrial NPP has nonetheless been influenced by changing climates, which in turn affects the magnitude of terrestrial carbon sinks. Since studies have suggested that NPP in the Northern Hemisphere has continued to increase over the past two decades (Nemani et al., 2003;

Zhao & Running, 2010), and because this hemisphere has been shown to be a $CO_2$ sink (Ciais et al., 1995) and, despite lessening impact in recent years (Yin et al., 2018), continues to act as one (Zhao & Running, 2010), an analysis of the carbon cycle in the Northern Hemisphere for recent decades is warranted. A study based upon simulations of carbon exchange has predicted that NPP will continue to increase in the 21st century (Cao & Woodward, 1998), and another simulation predicted that the terrestrial biosphere will act as a carbon sink until 2050 (Cox et al., 2000). However, the relation between increasing $CO_2$ and carbon cycle increases remains unquantified. This study aims to compare these two variables in each hemisphere to assist in determining the degree to which productivity may mitigate the continued rise of $CO_2$.

The results of Zhao and Running (2010) indicate that terrestrial NPP continued to increase in the Northern Hemisphere during the 2000s, following an increase which was seen in the 1980s and 1990s. While it has been suggested that NPP will continue to increase alongside $CO_2$ in the 21st century (Cao & Woodward, 1998; Cox et al., 2000), the degree to which the carbon cycle has increased with rising $CO_2$ has not yet been established. The purpose of the present study is to address this gap by evaluating whether $CO_2$ and the carbon cycle, defined here as the periodic component of $CO_2$ seen in Figure 1, have been increasing proportionally in recent decades, and if they have, to determine how the relative increases correspond. While correlation does not always imply causation, a correlation between the carbon cycle and atmospheric $CO_2$ would suggest the possibility of a causal relationship.

While previous studies on terrestrial NPP over large scales have used methods based on models (Nemani et al., 2003; Zhao & Running, 2010) or simulations (Cao & Woodward, 1998; Cox et al., 2000), the method for this study is based on an analysis of the measured atmospheric $CO_2$ concentrations directly via Fourier transforms. Fourier transforms provide accurate estimates of the underlying oscillatory components in the $CO_2$ measurements. The oscillating component with a period of 1 year corresponds to the annual carbon cycling as $CO_2$ is removed from the atmosphere through photosynthesis. This method will not serve as a direct measurement of NPP due to a potential lack of precise correspondence between actual NPP and the magnitude of the carbon cycle as determined by the method, but will instead analyze the magnitude of $CO_2$ oscillations, seen as the ripple in the curve of Figure 1, as a proxy for NPP. Consequently, this study will not be able to obtain magnitudes resulting specifically from terrestrial NPP, which was able to be differentiated from total NPP in the studies previously mentioned, and will instead analyze the carbon cycle resulting from atmospheric $CO_2$ in various latitude bands, a measurement which will likely be impacted by production in both terrestrial and oceanic environments.

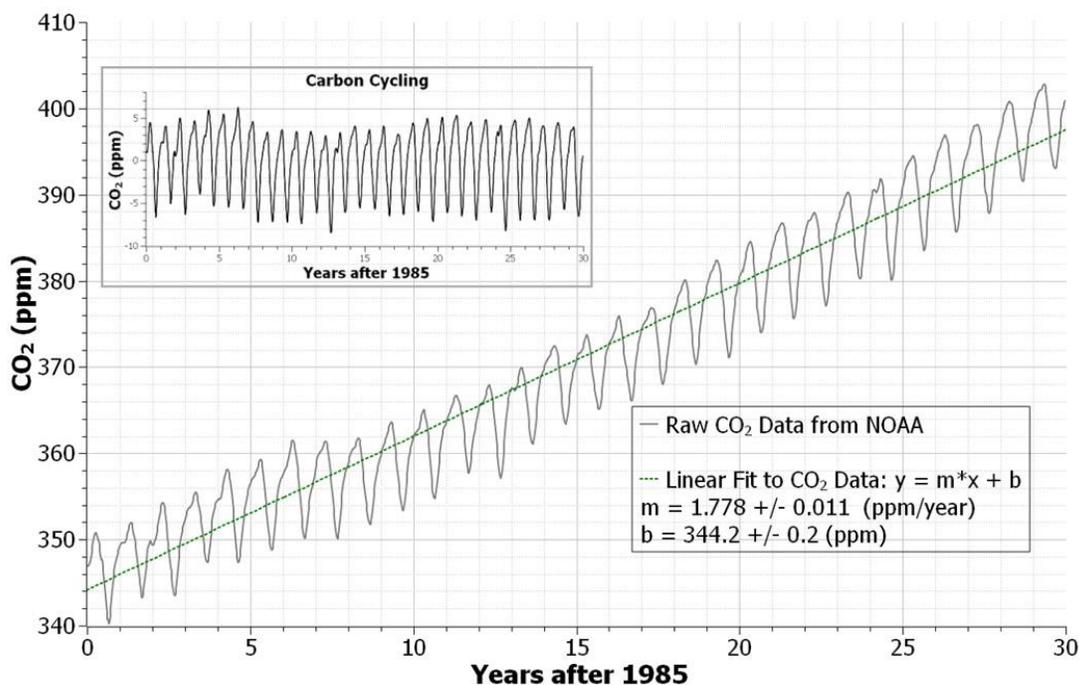

***Figure 1:*** *Raw $CO_2$ data from NOAA for latitudes 30° N to 40° N with linear fit. Carbon cycling, shown in the inset, is the periodic component of raw $CO_2$ data. Data for every distinct latitude band in 10° intervals was analyzed similarly.*

On the other hand, this measurement-based method is less likely to be limited by factors which can hamper studies based on algorithms or simulations, such as inaccurate modeling of values due to incorrect assumptions. Limitations of modeling approaches have impacted research in NPP trends, and the results of at least one of the studies (Zhao & Running, 2010) previously cited have been questioned due to a prediction of similar results by simple deduction from assumptions of the model used (Medlyn, 2011). The method of this study reduces the likelihood of such bias in results by having only one input (monthly averages of $CO_2$) instead of several parameters which could each add uncertainties or rely on additional assumptions, such as values sensitive to temperature in previous studies. The input was only subject to several straightforward mathematical processes, such as best-fitting to a polynomial and Fourier analysis. Due to the simplicity of prerequisite data and materials, the method of the study can also be repeated readily for multiple sets of $CO_2$ data at various sampling locations. Thus, this study's analysis of the data offers a complementary approach to existing models.

The hypothesis of this study was that as atmospheric $CO_2$ concentration increased, the carbon cycle would increase in a directly proportional fashion. It was hypothesized that the relative increases would correspond with a proportionality constant of 1, e.g., a 5% increase in atmospheric $CO_2$ would correspond to a 5% increase in the carbon cycled per year. A hypothesized proportionality constant of 1 served as a clear metric to determine whether carbon cycling was keeping pace with carbon cycling, because if the proportionality constant did not equal one, carbon cycling increases would be either falling behind or exceeding $CO_2$ increases.

**Method**

Datasets of monthly averages for $CO_2$ were gathered from the NOAA ESRL's Greenhouse Gas Marine Boundary Layer Reference (NOAA ESRL, 2017) for each 10° latitude band from 90° S to 90° N, making for a total of 18 datasets.

To compute the magnitude of the carbon cycle for a set of months, Fourier analysis was used on the periodic component of the $CO_2$ data for the respective months. In a previous study, Fourier analysis has been used on $CO_2$ data to separate seasonal oscillations from long term increase (Keeling et al., 1976), which indicates that the method could be utilized for an analysis of the seasonal variation itself. Because the seasonal variation, or periodic component, in $CO_2$ corresponds to the activity of plants and other photosynthetic organisms (Monroe, 2013), it can be inferred that the amplitude of this periodic component corresponds to a reasonable proxy for annual carbon cycling caused by producers. Therefore, the method of Fourier analysis was expected to accurately estimate magnitudes of yearly carbon cycling.

The method used for the Fourier analysis preparations followed a method described in a previous study on Fourier transforms (E. Courtney & M. Courtney, 2015) which was used to analyze a $CO_2$ dataset from Mauna Loa, Hawaii. First, data for missing months was filled using linear interpolation. In this present study, this step was completed in the preparation of the datasets by NOAA. Next, the full dataset from 1985 to 2015 for each of the 18 latitude bands was fit to a cubic polynomial, which was then subtracted from the dataset, leaving the periodic component of $CO_2$. An example of this periodic component for one latitude band is shown in the inset of Figure 1 and entitled "Carbon Cycling", because it is the proxy for carbon cycling in this study. These steps are in accordance with past studies on Fourier analysis of $CO_2$ data (Keeling et al., 1976), and fitting to a cubic polynomial is standard (Tans et al., 1990; Oltmans et al., 2006) and necessary to remove the large non-oscillatory component of the globally increasing $CO_2$ background, which allowed the subsequent Fourier analysis methods to more accurately determine the frequencies, phases, and magnitudes of the oscillatory components. The set was then divided into ten year intervals spaced every five years, from 1985 to 2015. This division process makes for a total of five intervals per dataset.

After the dataset was prepared in the above manner, Fourier analysis was performed on each ten year division using explicit integration software which has been shown to yield higher accuracy than fast Fourier transforms (E. Courtney & M. Courtney, 2015). With the outputs from this transform, amplitude peaks for frequencies 1/year, 2/year, 3/year, and 4/year were then recorded for each ten year interval. Figure 2 shows the Fourier transforms resulting from the Fourier analysis of the set for latitudes 30° N to 40° N.

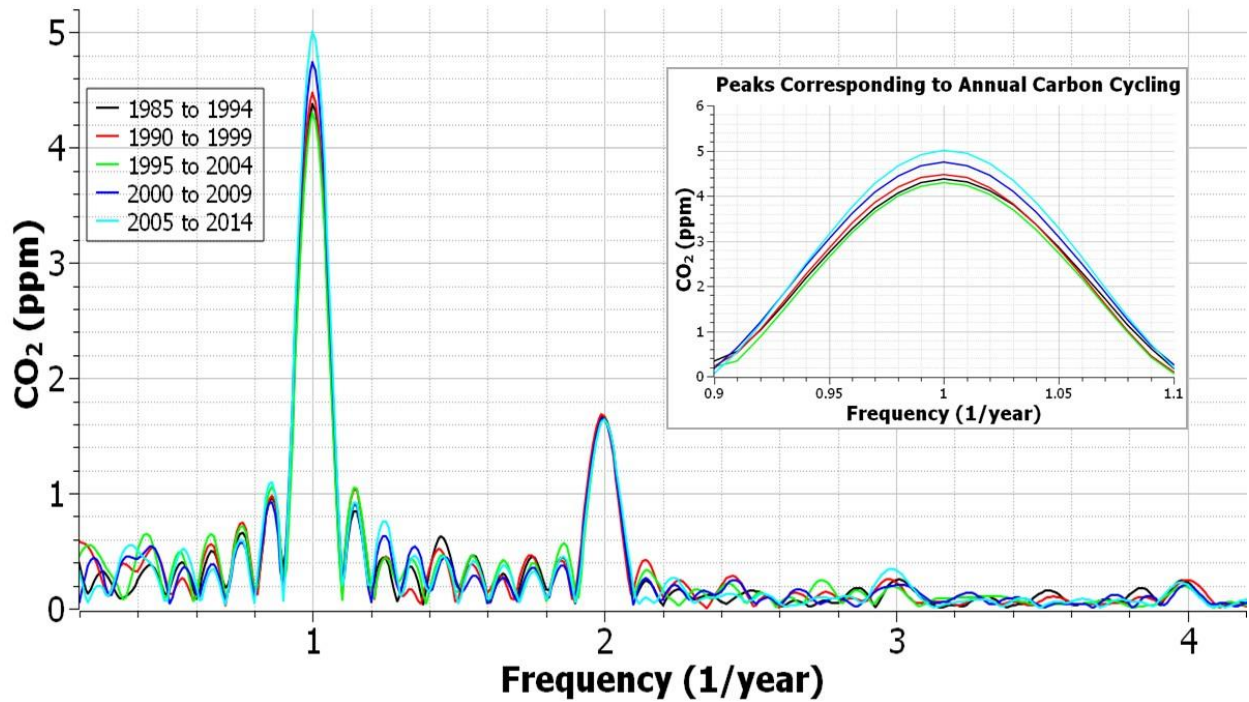

***Figure 2:*** *Fourier transforms resulting from Fourier analysis of the latitude band 30º N to 40º N split into staggered decades. The frequency of 1/year, or annual carbon cycling, has the greatest magnitude, which is detailed by staggered decades in the inset.*

The height of each peak with a frequency of once per year corresponds to the magnitude of the annual $CO_2$ oscillation over the corresponding time span. For each ten year interval, the height of each peak was recorded in a table alongside the middle year of the interval. For the set of peaks for each latitude band, a least-squares fit to a line was performed to determine the slope of the amplitude increase over time. This slope represents the annual increase in the $CO_2$ oscillation in ppm/year of $CO_2$.

Using the slope and the mean of resultant amplitude peaks for frequency 1/year, relative increase (expressed as a percent) was computed for each latitude band. This relative increase is given by $P_A = \frac{I_A}{M_A} \times 100\%$, where $I_A$ is the increase per decade of the amplitudes determined by a linear fit to the set of amplitudes, $M_A$ is the mean of the amplitudes, and $P_A$ is the relative increase of the amplitudes over the interval.

A least-squares fit and mean were also computed for raw $CO_2$ data for each latitude band. Relative increases for $CO_2$ were then computed using a similar formula, given by $P_{CO2} = \frac{I_{CO2}}{M_{CO2}} \times 100\%$, where $I_{CO2}$ is the increase per decade of $CO_2$ determined by a linear fit to the $CO_2$ concentration data, $M_{CO2}$ is the mean of the $CO_2$ data, and $P_{CO2}$ is the relative increase of the amplitudes over the interval.

For every latitude band, uncertainties were computed for each value using standard methods. Uncertainties in slopes of the best-fit lines were computed by regression software and multiplied by ten to give the uncertainty in the increase per decade. Uncertainties in the mean values were computed as the standard error of the mean. To calculate uncertainty in the relative increase, the error propagation formula

$$\Delta P = \Delta\left(\frac{I}{M}\right) = \left|\frac{\partial}{\partial m}\left(\frac{I}{M}\right)\right|\Delta M + \left|\frac{\partial}{\partial I}\left(\frac{I}{M}\right)\right|\Delta I = \left|\frac{I}{M^2}\right|\Delta M + \left|\frac{1}{M}\right|\Delta I$$

was used, where $\Delta P$ is the uncertainty in the relative increase, $I$ is the increase per decade, $M$ is the mean, $\Delta I$ is the uncertainty in the increase per decade, and $\Delta M$ is the uncertainty in the mean.

Means, increases per decade, and relative increases were also computed for both hemispheres and globally. First, data points for each decade for latitudes in the respective hemisphere – or all latitudes for the global computation - were first averaged to give the values of carbon cycling in the hemisphere or globe for that decade. Means, increases per decade, and relative increases were then found for these sets of averages using the same process described previously for individual latitude bands. Uncertainties were also computed for the values using the method described in the previous paragraph.

**Results**

To compare the rate of increase of carbon cycling with current $CO_2$ levels, relative changes in $CO_2$ were computed and plotted vs. latitude for the 18 latitude bands. Mean $CO_2$ concentrations and $CO_2$ increases were comparable across latitudes. Consequently, the relative increases were also similar for every latitude band, as shown in Figure 3.

For carbon cycling, the mean carbon cycling and increases per decade were computed for each latitude band to calculate the relative change for every latitude band. As seen in Figures 4 and 5, mean carbon cycling and increases in carbon cycling were lower in the Southern Hemisphere and higher in the Northern Hemisphere. Table 1 shows that the relative change in the Southern Hemisphere was slightly lower than the relative change in the Northern Hemisphere, but large uncertainties allow for the possibility that the values for each hemisphere were comparable.

In summary, relative increases in carbon cycling were lower than relative increases in $CO_2$ for most latitude bands, as seen in Figure 6. Additionally, relative increase in carbon cycling was lower than relative increase in $CO_2$ in the results for each hemisphere and globally, as seen in

Table 1. However, Figure 6 and Table 1 also show large error bars in carbon cycling relative increase, most of which have the values for relative increase in $CO_2$ within their range. These large uncertainties were also present in the global and hemispheric values, as noted previously.

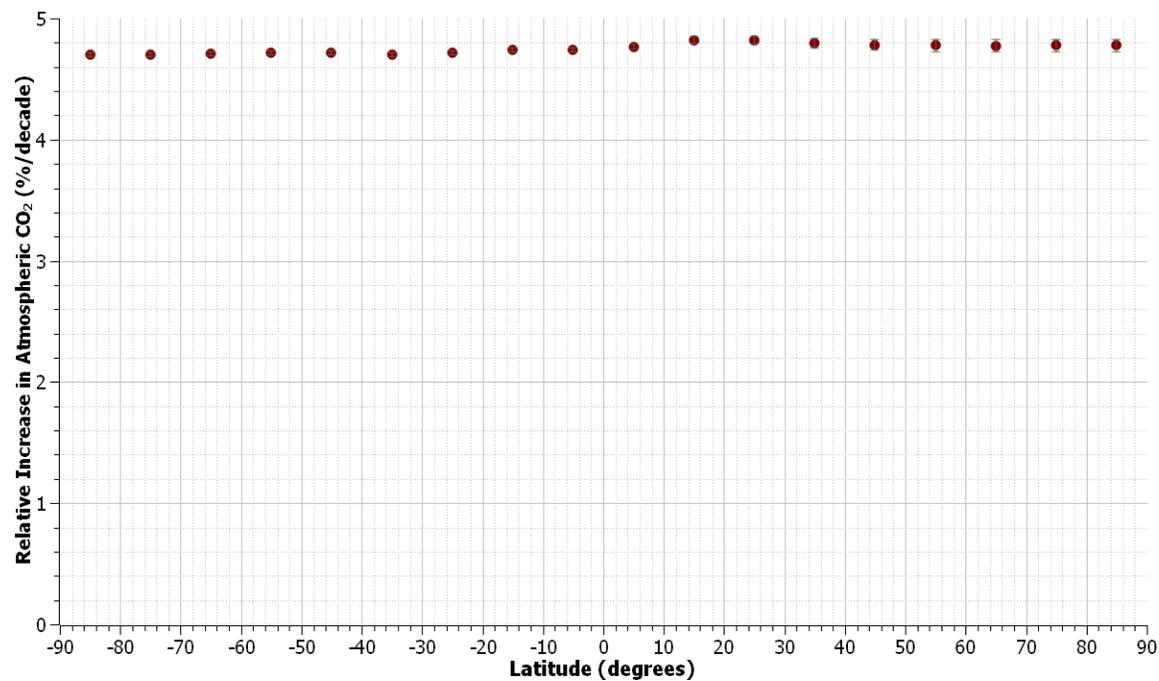

***Figure 3:*** *Relative increase in $CO_2$ for every latitude band. Across latitudes, the relative increase was also visually comparable and close to 4.7% for each band.*

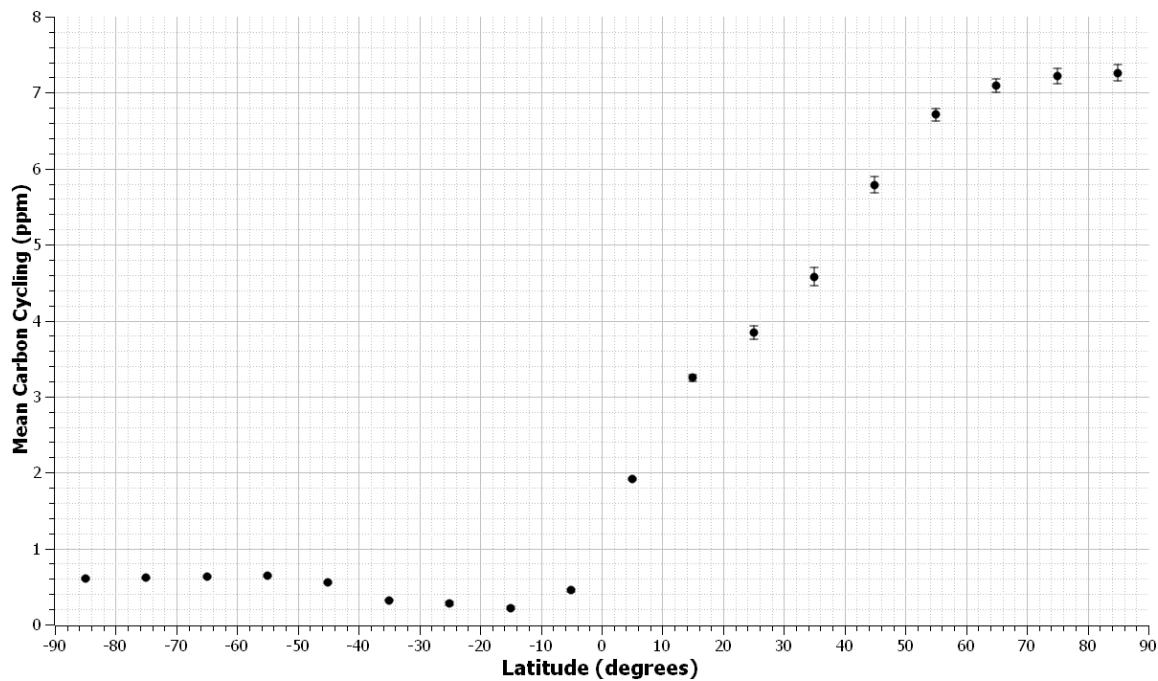

***Figure 4:*** *Mean magnitude of carbon cycling for each latitude band from 1985 through 2014, computed by averaging the carbon cycling of all decades in a given latitude band. The mean magnitude of carbon cycling was lower in Southern latitudes and higher in Northern latitudes.*

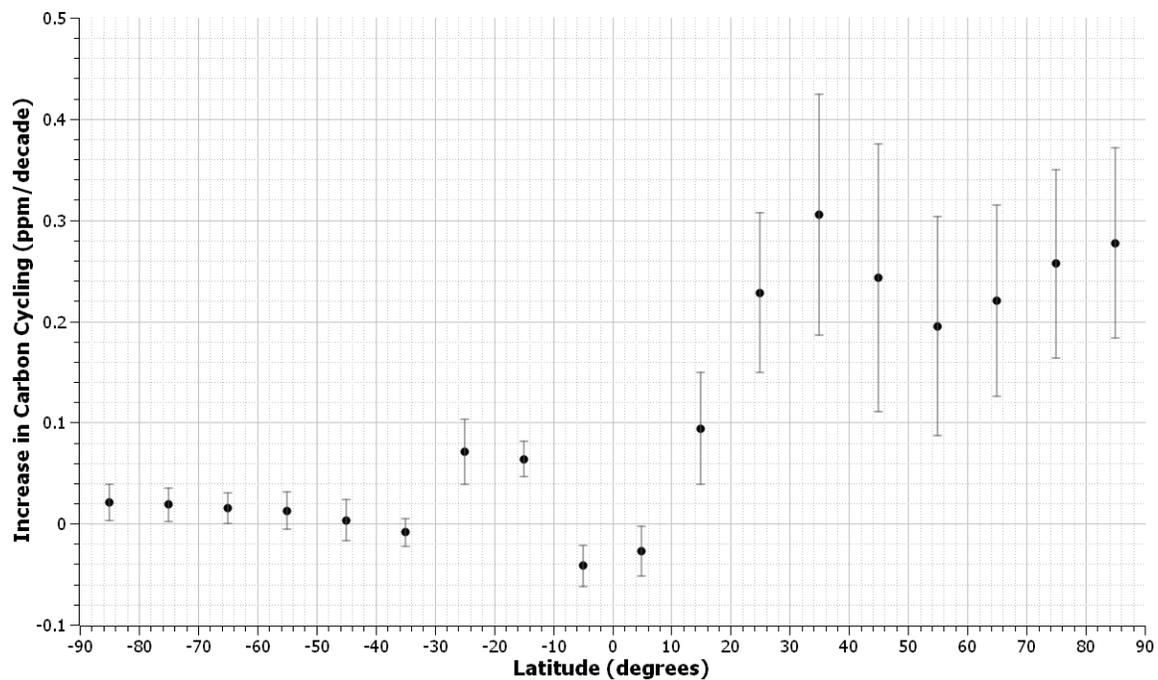

***Figure 5:*** *Increase in carbon cycling per decade for 1985 through 2014, which is ten times the slope of the best-fit line to the annual data points of carbon cycling for a given latitude band. Similar to mean carbon cycling, increase in carbon cycling was lower in Southern latitudes and higher in Northern latitudes.*

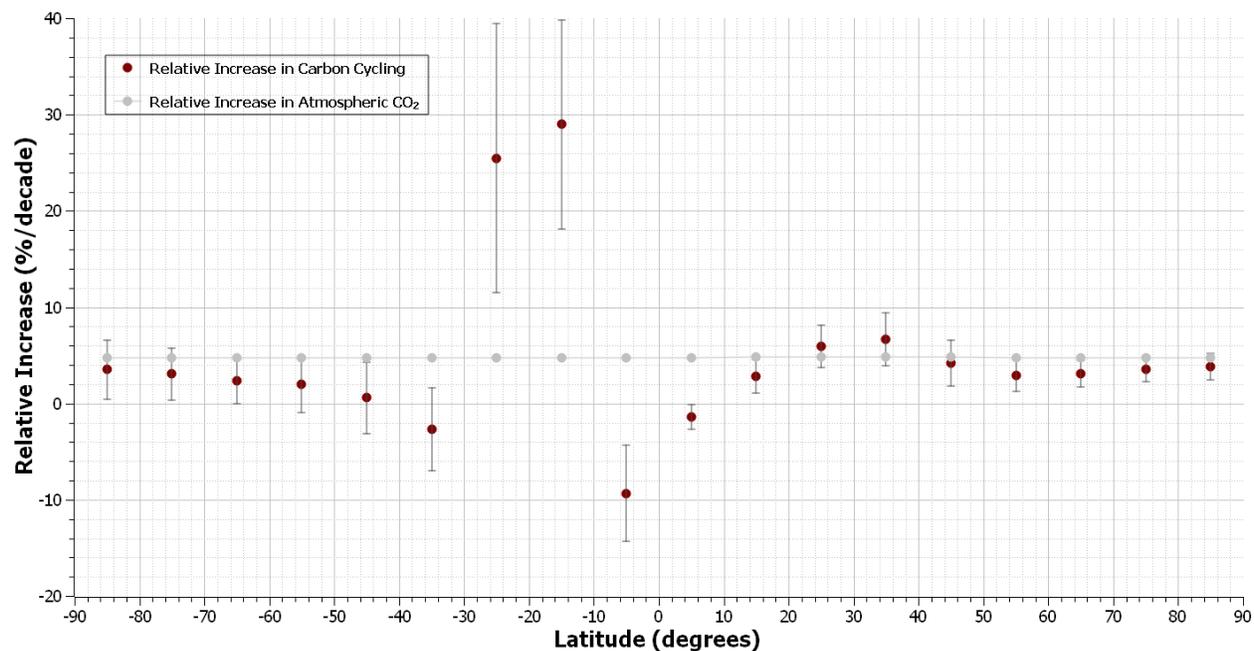

***Figure 6:*** *Relative change in carbon cycling for each latitude band, computed by dividing the increase in carbon cycling by the mean of carbon cycling for a given latitude band. Little distinction was observed between the relative changes of carbon cycling for latitudes in different hemispheres. Relative change in atmospheric $CO_2$, lightly shown, was greater than the relative change in carbon cycling for the majority of latitudes, but within the range given by most of the error bars.*

*Table 1:* *Means, increases, and relative increases in atmospheric $CO_2$ and carbon cycling for each hemisphere and globally, computed by averaging results from individual latitude bands. Values based on data from 1985 to 2014.*

| Set of Latitudes | Atmospheric $CO_2$ Mean (ppm) | Carbon Cycling Mean (ppm) | Atmospheric $CO_2$ Increase (ppm/decade) | Carbon Cycling Increase (ppm/decade) | Atmospheric $CO_2$ Relative Increase (%/decade) | Carbon Cycling Relative Increase (%/decade) |
|---|---|---|---|---|---|---|
| Northern Hemisphere | 371 | 5.30 $\pm$ 0.08 | 17.8 | 0.199 $\pm$ 0.084 | 4.79 $\pm$ 0.04 | 3.76 $\pm$ 1.63 |
| Southern Hemisphere | 368 | 0.479 $\pm$ 0.007 | 17.4 | 0.0173 $\pm$ 0.0085 | 4.72 $\pm$ 0.02 | 3.61 $\pm$ 1.83 |
| Global | 369 | 2.89 $\pm$ 0.04 | 17.6 | 0.108 $\pm$ 0.039 | 4.75 $\pm$ 0.02 | 3.75 $\pm$ 1.42 |

## Discussion and Conclusions

*Discussion of the Results and Comparison to Prior Studies*

Although the relative increases in carbon cycling globally and for both hemispheres indicate that carbon cycling is not keeping pace with atmospheric $CO_2$ increase, the uncertainties in the values are too large to draw this conclusion with statistical confidence. This was true for the majority of individual latitude bands as well, as seen in Figure 6.

Because of this result, the hypothesis that the ratio of carbon cycling relative increase to $CO_2$ relative increase would equal 1 was not supported. However, inspection of Table 1 shows that due to the conclusion's lack of statistical confidence, it is possible that the ratio could equal 1 for some latitude bands, thus supporting the hypothesis that carbon cycling is keeping pace with $CO_2$.

The global result of 3.75% ($\pm$ 1.42%) per decade increase in carbon cycling found in this study is consistent with the result from Nemani et al. (2003), which suggested that NPP increased 6% globally from 1982 through 1999, or around 3.3% per decade. This suggests carbon cycling is a suitable proxy for NPP. However, this study's result of a 4.75% ($\pm$ 0.02%) per decade increase in atmospheric $CO_2$ is not consistent with the increase of 9% from 1982 to 1999, or around 5.0% per decade, reported by the same study. This difference can be attributed to the other study's computation making use of only the endpoint years in their interval, while this study's computations are based on the best-fit line to all monthly $CO_2$ points in the interval.

The findings of Zhao & Running (2010) indicated a net decrease in global NPP during the decade of 2000 to 2009, which, on first inspection, contradicts the positive global result of carbon cycling in this study. However, the method of this study is not suitable for either supporting or contradicting this claim in that several decades were analyzed, while the previous study was based on data for a single decade. In addition, the finding of a net decrease in global NPP was suggested to be due to a large decrease in the NPP of the Amazon and other particular regions, which impacted only two or three latitude bands. The net change in NPP for the Northern Hemisphere was reported to be positive in the same study, which indicates this study and the 2010 study may have more consistent results than seen from the value corresponding to global change in NPP.

The result of positive relative increases in most latitudes also affirms predictions that NPP will continue to increase in the 21$^{st}$ century (Cao & Woodward, 1998; Cox et al., 2000). Future studies using a similar method could monitor this increase in NPP as more decades pass, allowing for stronger support or contradiction of the predictions previously made.

Although the large uncertainties in carbon cycling leave ambiguity in the answer to the primary question of whether carbon cycling is keeping pace with atmospheric $CO_2$, this study accomplished the goal of providing quantitative estimates for carbon cycling as a proxy for primary production, which paves the way for future studies examining the relationship between atmospheric $CO_2$ concentration and primary production.

*Discussion of Other Factors Potentially Impacting Carbon Cycling Increase*
Although this study compared increases in $CO_2$ and carbon cycling, the causation of the increase in carbon cycling was not examined, and thus $CO_2$ increase may not be the primary factor impacting carbon cycling on a global scale. One factor which could be affecting the relation between $CO_2$ and carbon cycling is deforestation, which would particularly affect latitude bands containing tropical forests, such as 0° S to 10° S. Studies based on climate models have indicated that deforestation would not only decrease carbon cycling in a particular region, but also that the destruction of the plants would release additional $CO_2$ into the atmosphere (Cramer et al., 2004; Bala et al., 2007). In combination, these results indicate that deforestation may cause a disparity between increases in carbon cycling and $CO_2$, hence blurring any possible causation between increased $CO_2$ and carbon cycling.

This study's results for the latitude band 0° S to 10° S are consistent with the possibility for deforestation affecting the results, with Figure 6 showing that this latitude band has the lowest relative increase in carbon cycling. However, relative increase in $CO_2$ from 0° S to 10° S is comparable with results for other latitude bands, so the disparity between the relative increases in this case is due to a low value of carbon cycling increase, not an additional increase in $CO_2$. Generalizations about deforestation should also not be implied from this single data point, and the case of 0° S to 10° S only highlights the possibility for the additional factor of deforestation affecting the results of this study.

Changes in climates may also affect carbon cycling increase. Zhao & Running (2010) claimed a decrease in NPP for 2000 to 2009 due to increased temperature in the tropics, while a past study (Melillo et al., 1993) indicated a more ambiguous relationship between changing climates and NPP. Thus, while the results of this current study may have been affected by climatic factors, the exact nature of this effect is difficult to determine given both the scope of this study and the ambiguous earlier work on the relationship between climate change and NPP increase. Additionally, climatic factors may change from decade to decade, making the method of this study, based on data from a range of decades, unsuitable for accurately measuring these impacts.

*Possible Future Work and Applications*
As data sets from individual monitoring sites are also available from NOAA, a similar study performed on these individual sites has the potential to provide results for comparison to the latitude band averaged data analyzed in this study. At the current time, many individual monitoring sites have few decades of data, and sites with twenty five or more years of data are limited primarily to North America. In the next decade, however, more sites across the globe will likely have data available for at least twenty five years, which could be a suitable basis for future studies.

Since increases in $CO_2$ have implications which are not limited to climate science, ranging from impacts in communities of primary producers (Low-Decarie, Fussmann, & Bell, 2014) to applications in crop science (Sakurai et al., 2014), potential applications of this analysis of the carbon cycle include these areas and others.

**Acknowledgements**

**Competing Interests**
The authors declare no competing interests.